\documentclass[epj]{svjour}
\usepackage{graphicx}
\usepackage{amsmath,amssymb}
\begin{document}
\title{Influence of bulk mass distribution on orbital precession of S2 star in Yukawa gravity}
\author{Predrag Jovanovi\'{c}$^1$\thanks{\emph{e-mail:} pjovanovic@aob.rs}, Du\v{s}ko Borka$^2$, Vesna Borka Jovanovi\'{c}$^2$ \and Alexander F. Zakharov$^{3,4,5}$}
\authorrunning{Jovanovi\'{c} et al.}
\institute{$^1$Astronomical Observatory, Volgina 7, P.O. Box 74, 11060 Belgrade, Serbia \\
$^2$Department of Theoretical Physics and Condensed Matter Physics (020), Vin\v{c}a Institute of Nuclear Sciences - National Institute of the Republic of Serbia, University of Belgrade, P.O. Box 522, 11001 Belgrade, Serbia \\
$^3$Institute of Theoretical and Experimental Physics, 117259 Moscow, Russia \\
$^4$Bogoliubov Laboratory for Theoretical Physics, JINR, 141980 Dubna, Russia \\
$^5$National Research Nuclear University MEPhI (Moscow Engineering Physics Institute), 115409, Moscow, Russia}
\date{Received: date / Revised version: date}
%

\abstract{In this study we investigate possible applications of observed S2 orbit around Galactic Center for constraining the Yukawa gravity at scales in the range between several tens and several thousands astronomical units (AU) to obtain graviton mass constraints. In our model we suppose that bulk distribution of matter (includes stellar cluster, interstellar gas distribution and dark matter) exists near Supermassive Black Hole (SMBH) in our Galactic Center. We obtain the values of orbital precession angle for different values of mass density of matter and we require that the value of orbital precession is the same like in General Relativity (GR). From that request we determine gravity parameter $\lambda$ and the upper value for graviton mass. We found that in the cases where the density of extended mass is higher, the maximum allowed value for parameter $\lambda$ is smaller and the upper limit for graviton mass is higher. It is due to the fact that the extended mass causes the retrograde orbital precession. We believe that this study is a very efficient tool to evaluate a gravitational potential at the Galactic Center, parameter $\lambda$ of the Yukawa gravity model, and to constrain the graviton mass.}
	
%
\PACS{
      {}{Modified theories of gravity} \and {}{Black hole physics} \and {}{Galactic Center}
     } 
\maketitle

\section{Introduction}
\label{sec01}

One of the most important problems to solve in modern theoretical physics are the Dark matter (DM) \cite{zwic33} and Dark Energy (DE) \cite{turn99} problems. These problems are fundamental and difficult for the conventional GR approach for gravity \cite{zakh09,wein08}. One of the possible ways to solve DM and DE puzzle is to change the gravity law as it was done for DM in \cite{milg83,beke04} and for both DM and DE problems with the $f(R)$ approach \cite{capo02,carr04,capo06}. We investigate properties of the extended theories of gravity \cite{zakh06,capo11,capo11b,bork12,bork16} with aim to extend the positive results of GR and, on the other hand, to cure its shortcomings. There is an opinion that an introduction of alternative theories of gravity (in our case Extended theories of gravity) could give explanation of observational astronomical data without including DM and DE problems. Besides above mentioned fundamental issues, like dark energy and dark matter, these theories can help us to constrain graviton mass by comparing their predictions with precise astronomical observations of the orbit of S2 star near SMBH in our Galactic Center \cite{genz03,wein05,ghez08,gill09a,gill09b,genz10,hees17,chu18,hees20}\footnote{The remarkable studies got the high recognition in scientific community and Reinhard Genzel (VLT) and Andrea Ghez (Keck) were awarded the Nobel prize in physics in 2020.}.

The B-type main-sequence star S2 is a massive young star of $\approx 15\ M_\odot$ ($M_\odot$ being the solar mass) which is one of the brightest members of the so-called S-star cluster orbiting the central SMBH of the Milky Way. Members of this stellar cluster and their motions around the SMBH are monitored for almost 30 years by large observational facilities, such as New Technology Telescope and Very Large Telescope in Chile, as well as Keck telescope in Hawaii, USA, primarily in the near-infrared K spectral band, due to interstellar matter which limits their visibility in the optical wavelengths.

Also, an experimental detection of graviton is a very hard problem to solve and there are different ways to evaluate a graviton mass if it is non-vanishing \cite{zakh16,zakh18a,zakh18b,zakh20}. We use Yukawa gravity, one among the gravity theories with non-vanishing graviton mass \cite{zakh16,zakh18a,bork13} to give constraint of graviton mass. In the few recent publications reporting about the discovery of gravitational waves from the binary black hole system, the LIGO-Virgo collaboration obtained the graviton mass constraints \cite{abbo16,abbo17a,abbo17b,abbo17c} and in the last years the constraint was significantly improved, in particular, based on a joint analysis of events from the first (O1) and the second (O2) observing runs or in other words, the events were collected in the first LIGO--Virgo Gravitational-Wave Transient Catalog (GWTC-1), the authors found that graviton mass should be $m_g < 4.7\times 10^{-23}$~eV \cite{abbo19}, while adding events from the first part of the third observational run (O3a) to GWTC-1 to form the second LIGO--Virgo Gravitational-Wave Transient Catalog (GWTC-2), the authors found that $m_g < 1.76 \times 10^{-23}$~eV \cite{ligo20}. In our previous papers, constraint of graviton mass has been obtained from an analysis of trajectories of bright stars near the Galactic Center \cite{zakh16,zakh18a,zakh18b,zakh20,bork13} assuming a potential of bulk distribution of matter is negligible in comparison with a potential of a point like mass. In this paper we consider orbital S2 star precession due to an inclusion of potential of a bulk distribution of matter \cite{genz03,rubi01,nuci07,zakh07,pret09,zakh14,doku15,doku15b,amor19}.

In this paper we found constraints on parameters of Yukawa gravity (particular case of the so-called massive gravity theories) with request that the obtained values of orbital precession angle are the same like in GR, but for different values of mass density of matter. We use a weak field limit for Yukawa gravitation potential.

\begin{figure*}[ht!]
\centering
\includegraphics[width=0.51\textwidth]{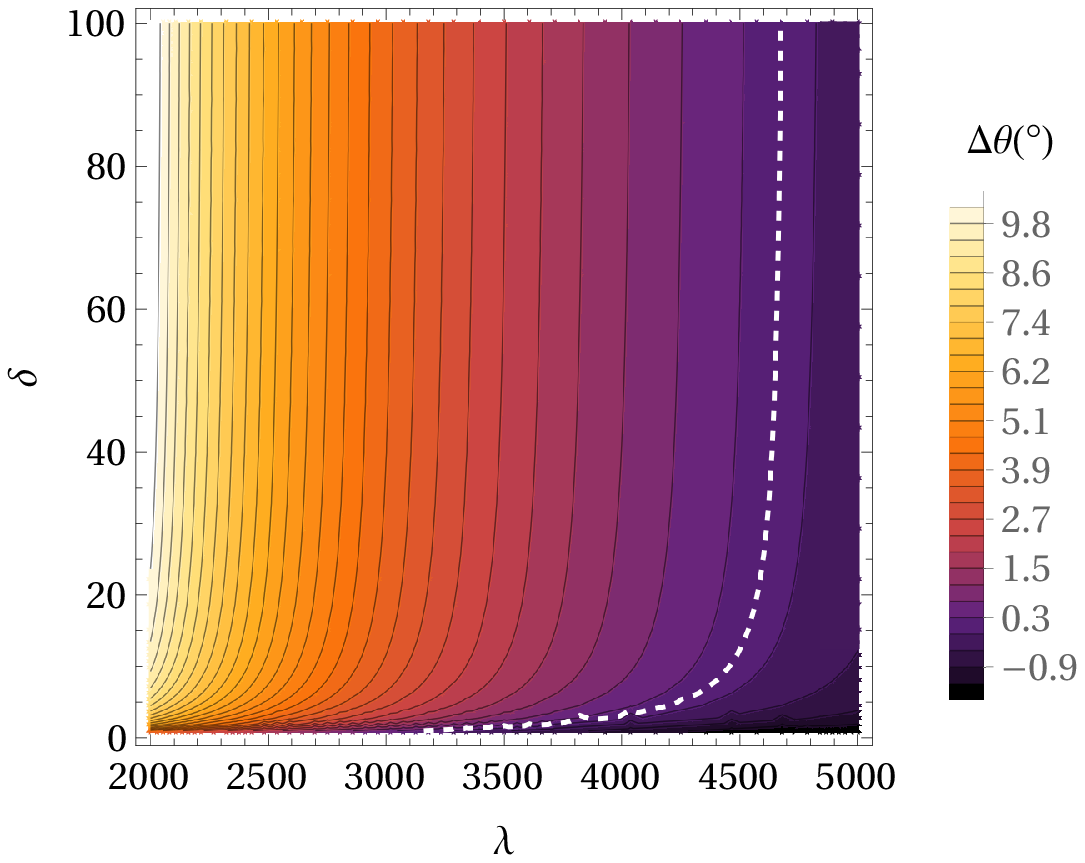}
\includegraphics[width=0.48\textwidth]{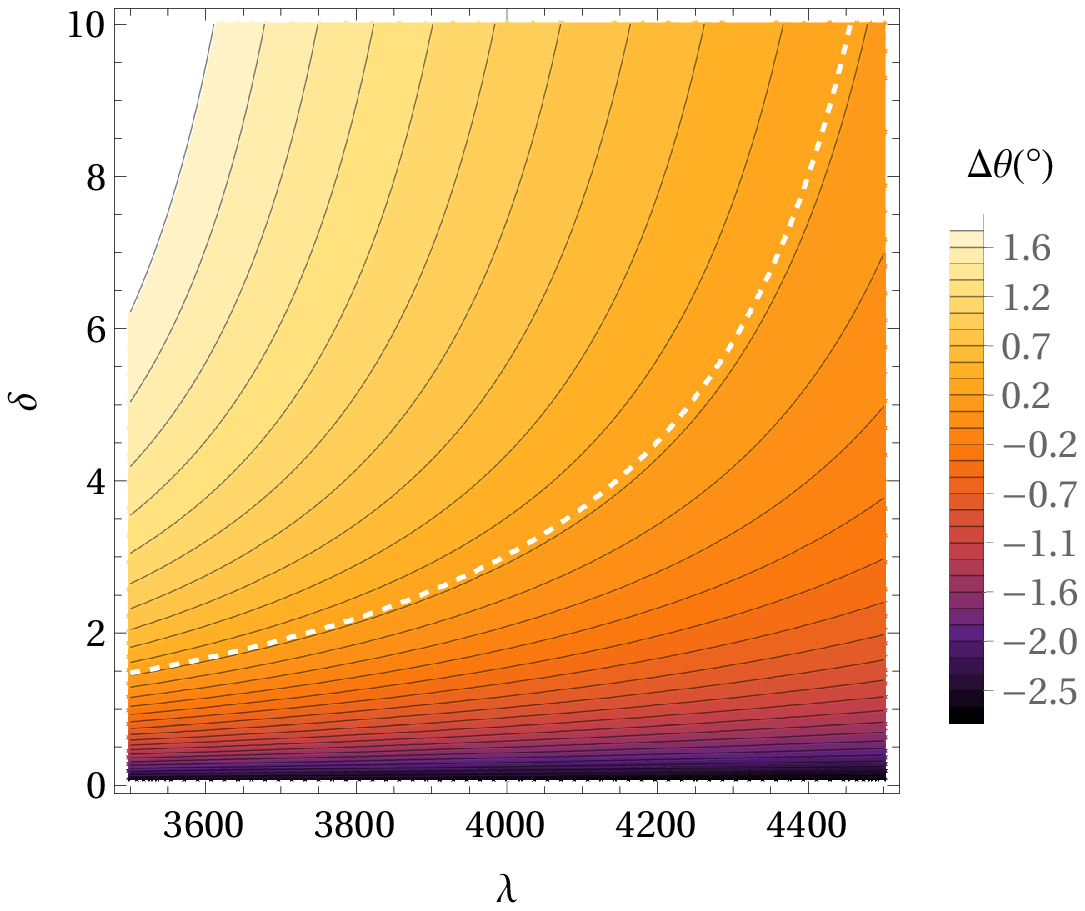}
\caption{The precession per orbital period for S2 star in ${\delta}-{\lambda}$ parameter space in the case of Yukawa modified gravity potential with extended mass distribution. The mass density distribution of extended matter is $\rho_0$ = $2 \times 10^8 M_\odot \mathrm{pc^{-3}}$. Right panel represents enlarged part of the left panel where the obtained values of orbital precession angle are very close to the value for precession angle in GR ($0^\circ.18$) which is designated by dashed line. With a decreasing value of precession angle  colors are darker. Parameter ${\lambda}$ is expressed in AU.}
\label{fig01}
\end{figure*}

\begin{figure*}[ht!]
\centering
\includegraphics[width=0.51\textwidth]{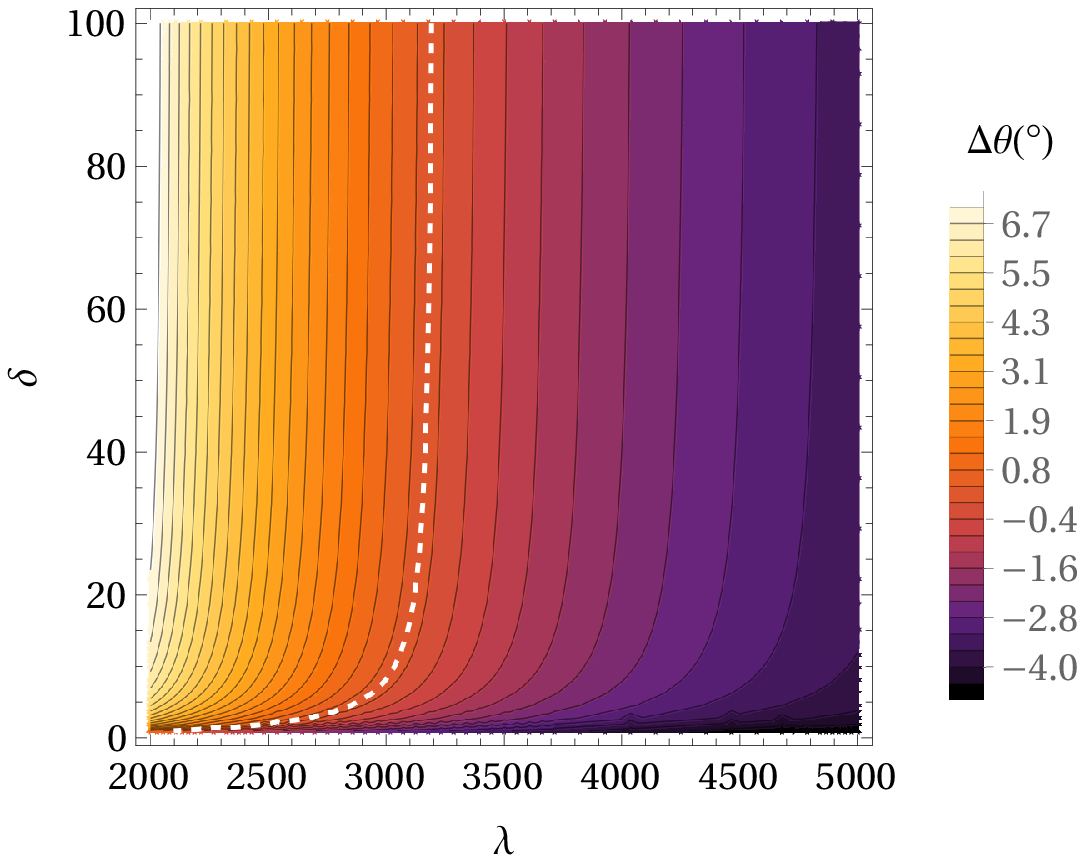}
\includegraphics[width=0.48\textwidth]{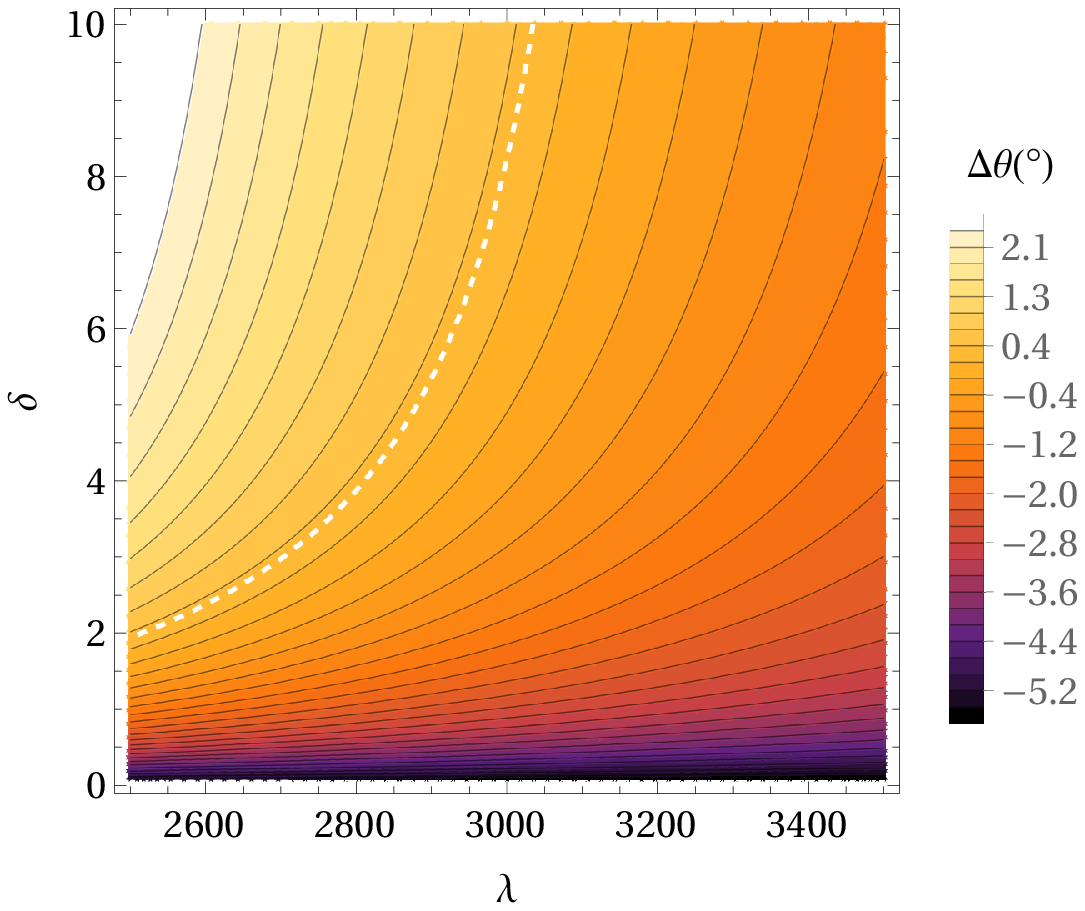}
\caption{The same as in Fig. \ref{fig01}, but for the value of the mass density distribution of extended matter $\rho_0$ = $4 \times 10^8 M_\odot \mathrm{pc^{-3}}$.}
\label{fig02}
\end{figure*}

\begin{figure*}[ht!]
\centering
\includegraphics[width=0.51\textwidth]{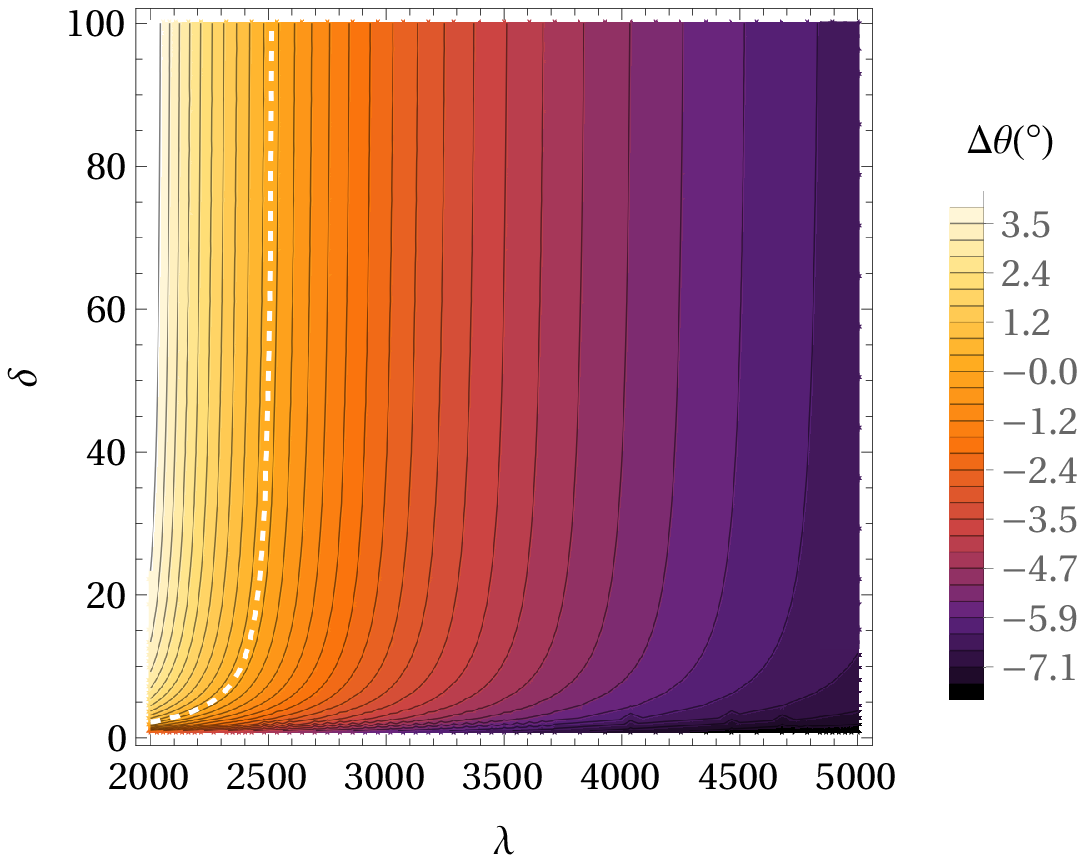}
\includegraphics[width=0.48\textwidth]{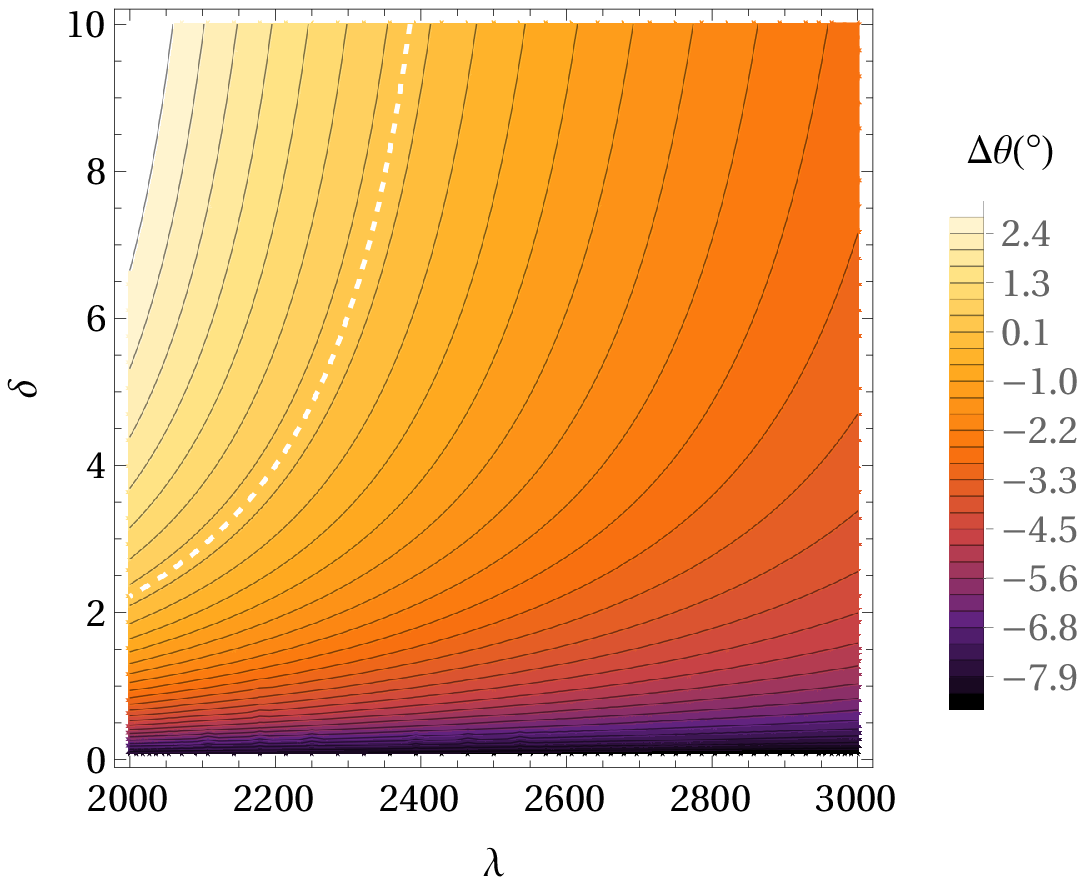}
\caption{The same as in Figs. \ref{fig01} and \ref{fig02}, but for the value of the mass density distribution of extended matter $\rho_0$ = $6 \times 10^8 M_\odot \mathrm{pc^{-3}}$.}
\label{fig03}
\end{figure*}

\begin{figure*}[ht!]
\centering
\includegraphics[width=0.51\textwidth]{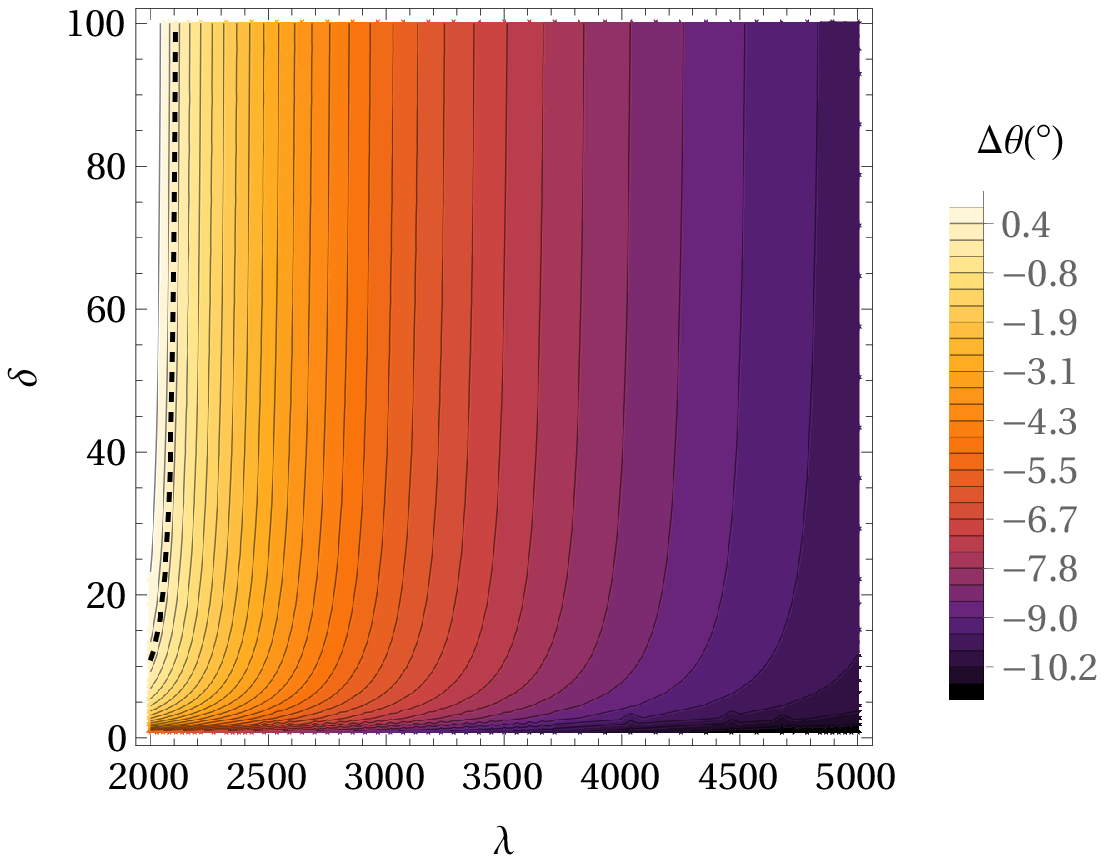}
\includegraphics[width=0.48\textwidth]{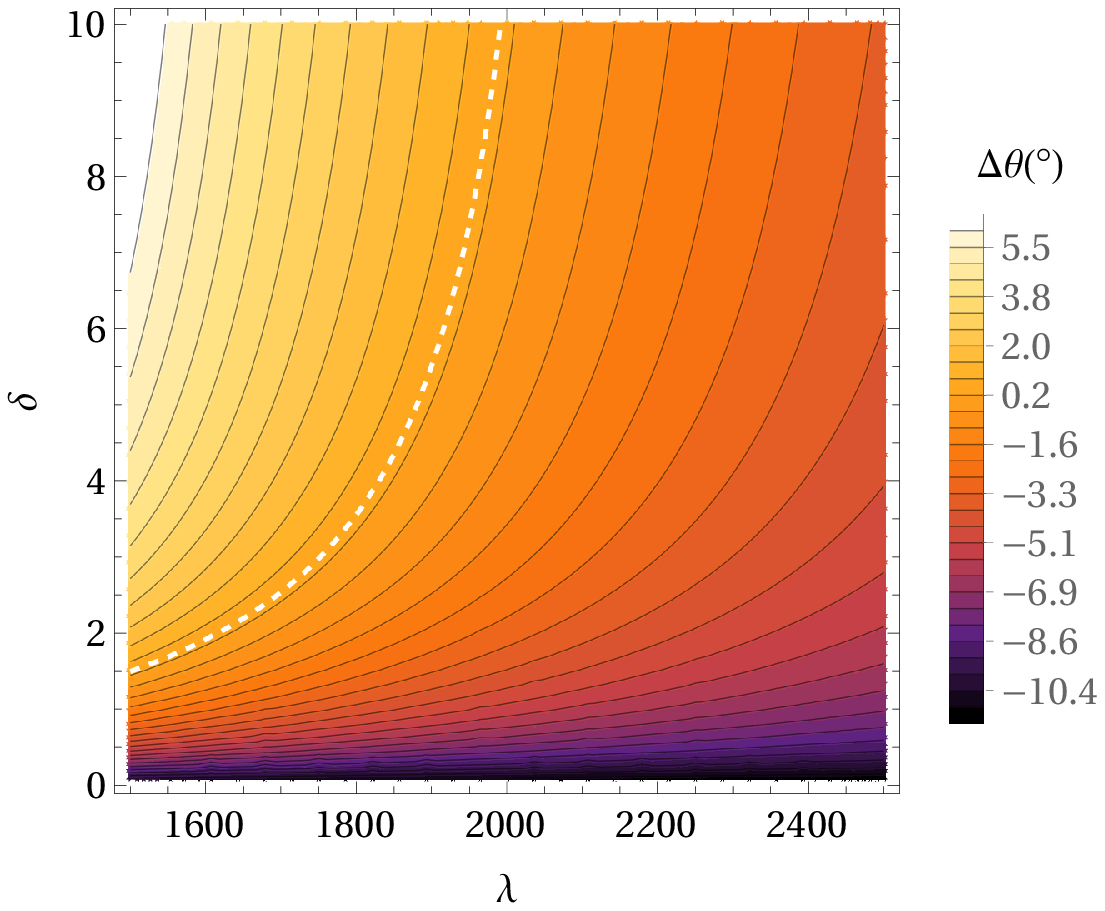}
\caption{The same as in Figs. \ref{fig01}, \ref{fig02} and \ref{fig03}, but for the value of the mass density distribution of extended matter  $\rho_0$ = $8 \times 10^8 M_\odot \mathrm{pc^{-3}}$.}
\label{fig04}
\end{figure*}

\begin{figure*}[ht!]
\centering
\includegraphics[width=0.59\textwidth]{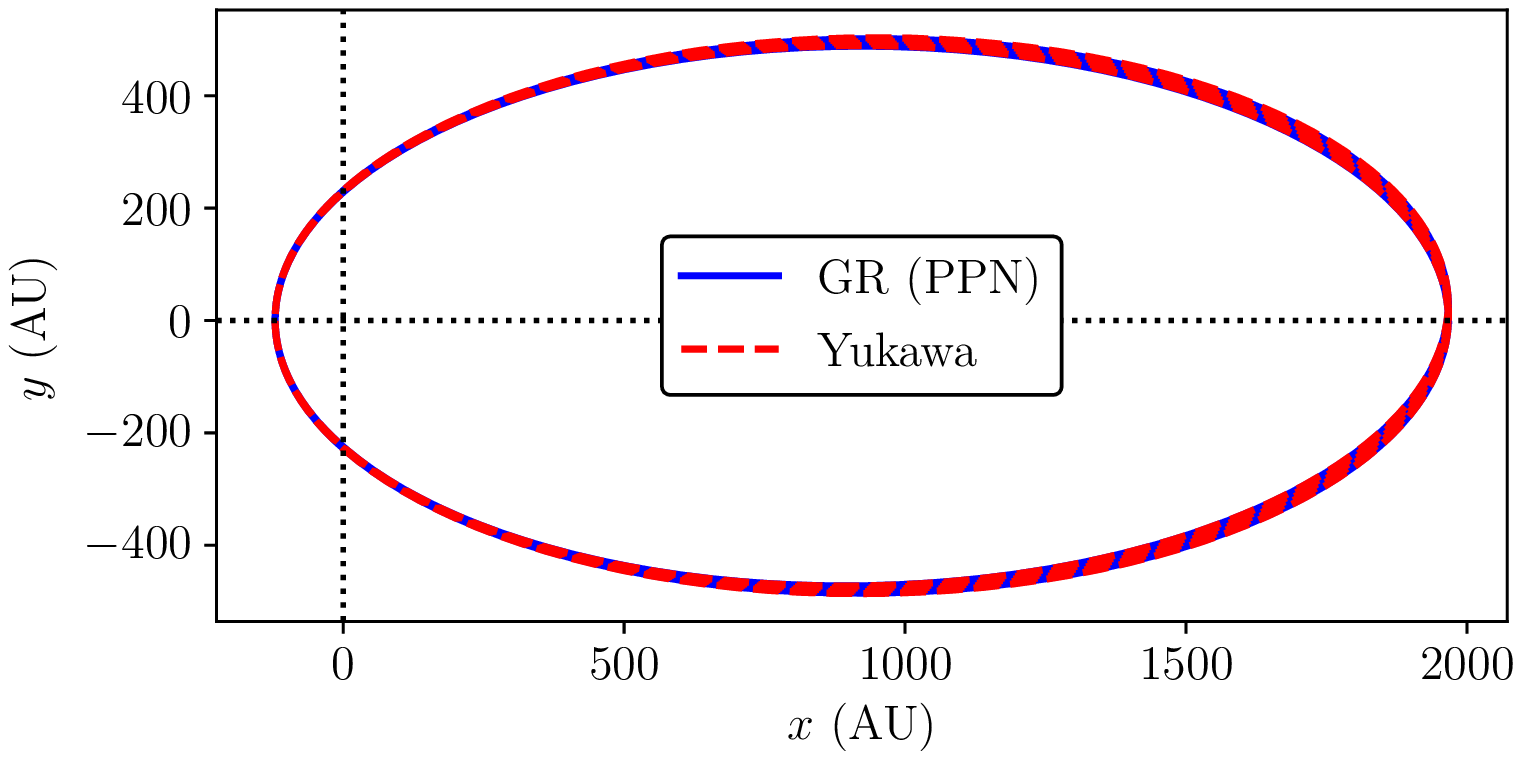}\hfill
\includegraphics[width=0.41\textwidth]{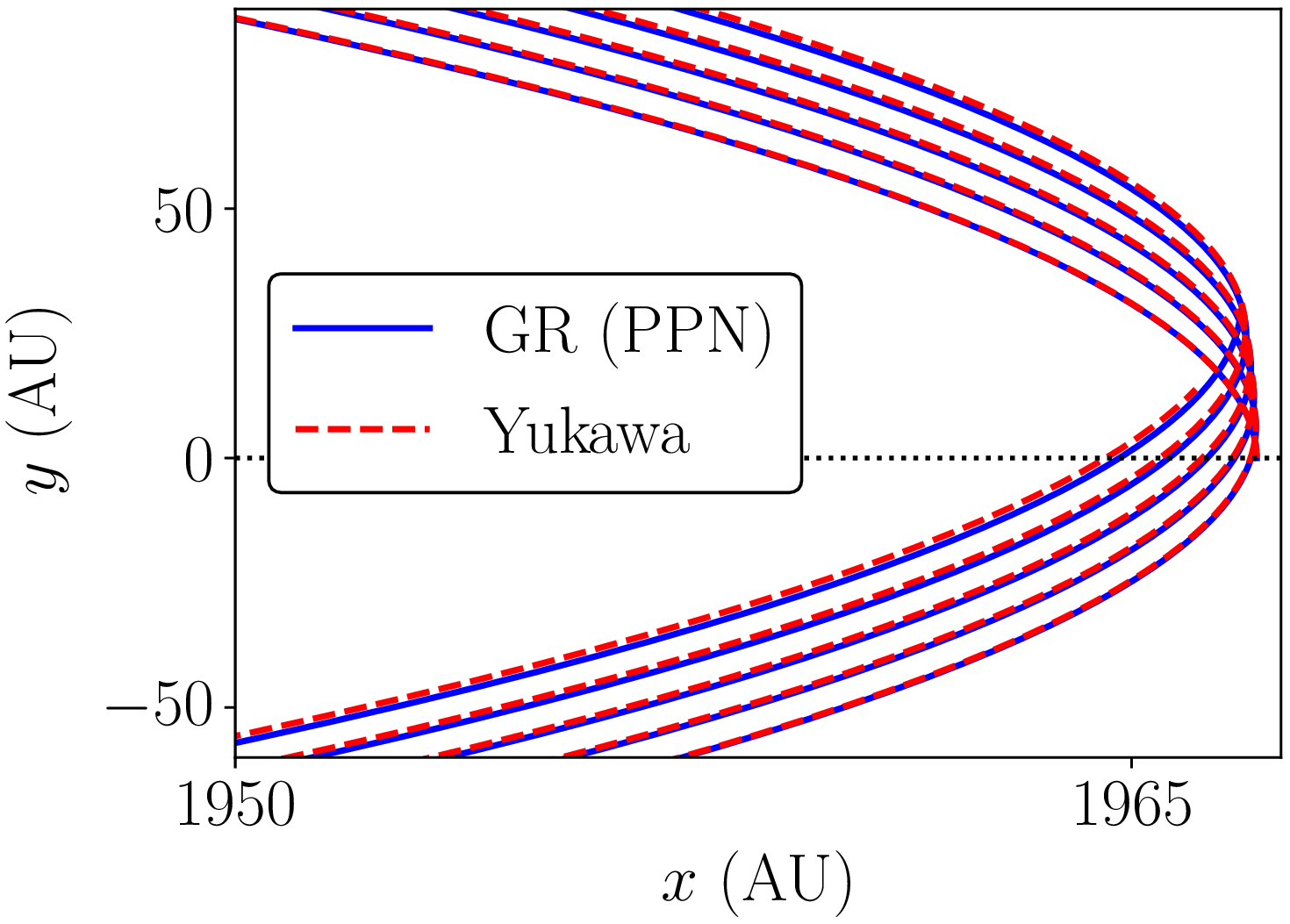}
\caption{Comparison between the simulated orbits of S2-star in GR (blue solid line) and in Yukawa gravity with the mass density distribution of extended matter $\rho_0$ = $2 \times 10^8 M_\odot \mathrm{pc^{-3}}$ (red dashed line) during five orbital periods. Region around apocenter is zoomed in the right panel in order that small orbital precession of $\Delta\varphi=722.1''= 0^\circ.2$ is visible. The values of parameters are $\lambda$ = 3130 AU and $\delta$ = 1.}
\label{fig05}
\end{figure*}

\begin{figure*}[ht!]
\centering
\includegraphics[width=0.59\textwidth]{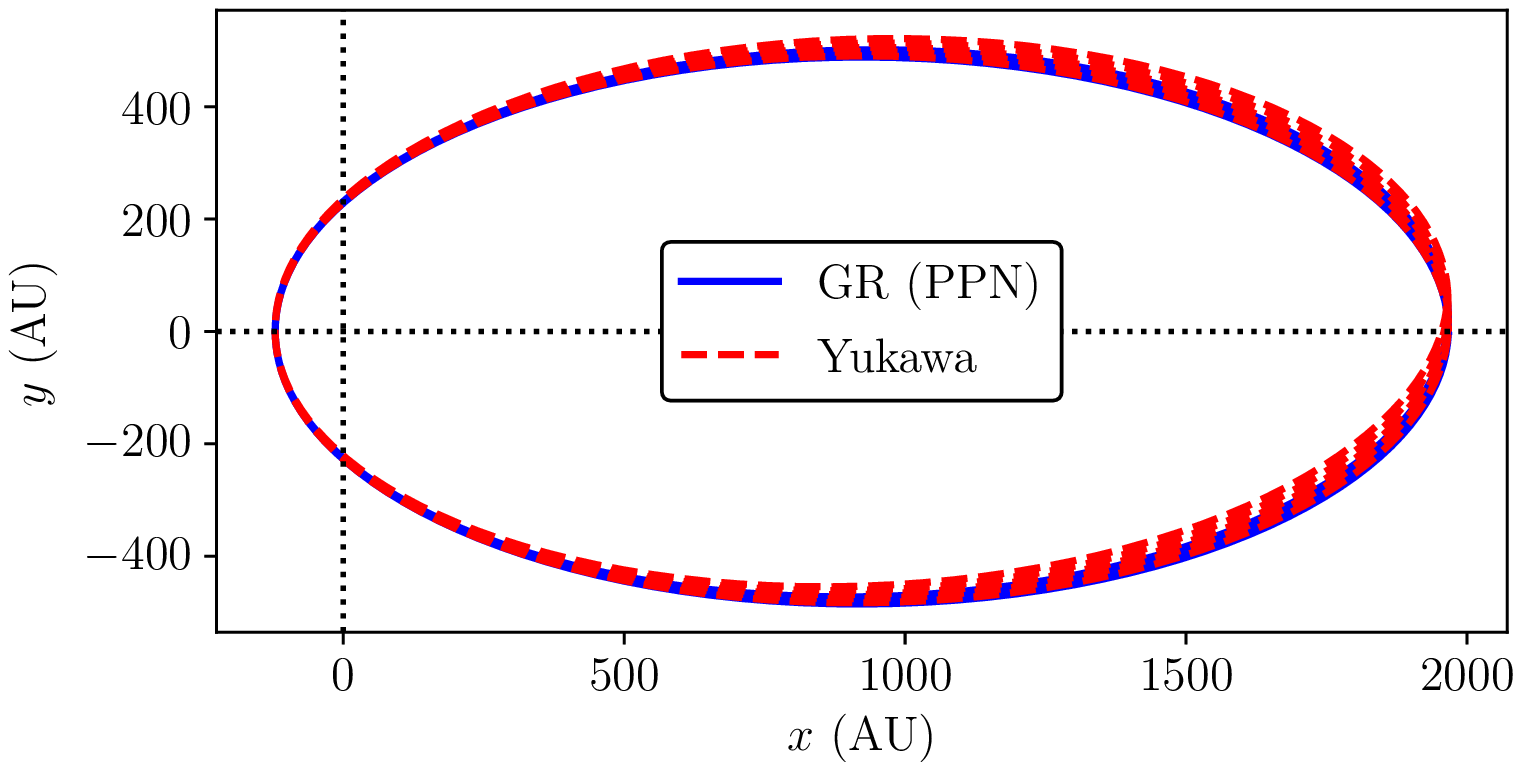}\hfill
\includegraphics[width=0.41\textwidth]{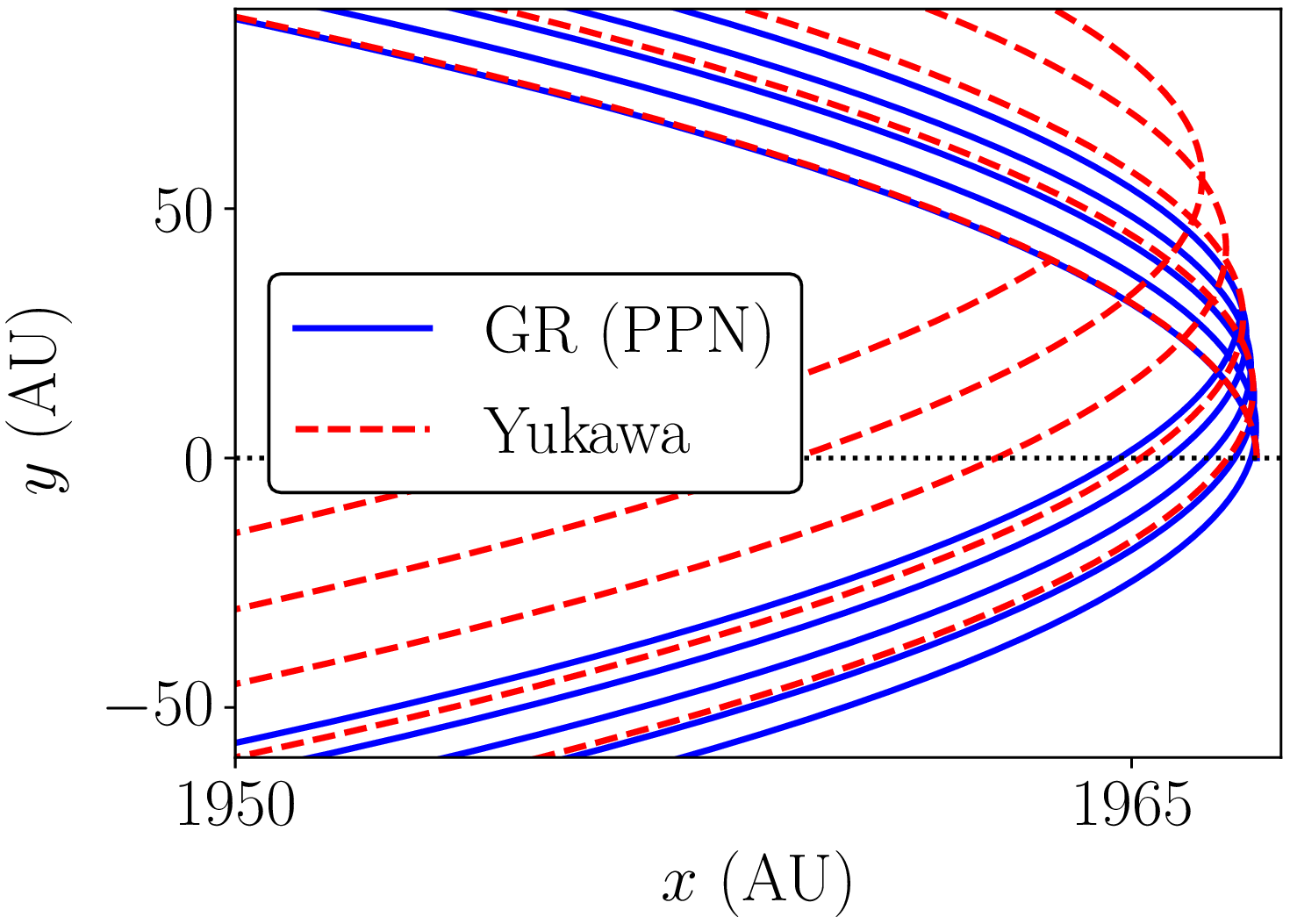}
\caption{The same as in Fig. \ref{fig05}, but for the following values of parameters: $\lambda$ = 3000 AU and $\delta$ = 1.}
\label{fig06}
\end{figure*}

\section{Orbital precession in Yukawa gravity}
\label{sec02}

Orbital precession of investigated star (in our case S2 star) is influenced by other stars, gas, and dark matter. Stars are expected to be the dominant component of the extended galactic mass distribution near the central SMBH. In order to investigate orbital precession of S2 star we assumed the following gravitational potential \cite{bork13}:

\begin{equation}
\Phi_Y(r)=-\dfrac{GM}{(1+\delta)r}\left[{1+\delta e^{-\dfrac{r}{\Lambda}}} \right],
\label{equ01}
\end{equation}

\noindent where $\Lambda$ is the range of Yukawa interaction and $\delta$ is a universal constant \cite{capo09b,card11}.

Also, we assume a bulk distribution of mass $M(r)$ in the central regions of our Galaxy.

\begin{equation}
M(r)=M_{BH}+M_{ext}(r),
\label{equ02}
\end{equation}
which is constituted by the central black hole of mass $M_{BH}= 4.3 \times10^6 M_\odot$ \cite{gill09a} and an extended distribution of matter with total mass $M_{ext}(r)$. Bulk extended mass distribution include a stellar cluster, interstellar gas and dark matter contained within some radius $r$. For the mass density distribution of extended matter we adopted a double power-law mass density profile \cite{genz03,pret09,amor19}:
\begin{equation}
\rho(r)=\rho_0\left( \frac{r}{r_0}\right) ^{-\alpha },\;\alpha
=\left\{
\begin{array}{ll}
2.0\pm 0.1, & r\geq r_0 \\
1.4\pm 0.1, & r < r_0
\end{array}
\right.
\label{equ03}
\end{equation}

\noindent where $\rho_0$ is varied from 2 to 8 $\times 10^{6}\ M_{\odot }\,\mathrm{pc}^{-3}$ and $r_0=10^{\prime\prime}$.

This broken power law model can be translated into a single power-law model throughout the whole region of special interest for us (orbit of S2 star). Therefore, we can choose only one value for $\alpha$ = 1.4.

Using above mentioned formulas, we get the following expression for the extended mass distribution:
\begin{equation}
M_{ext}(r)=\frac{4\pi\rho_0 r_0^\alpha}{3-\alpha}r^{3-\alpha}.
\label{equ04}
\end{equation}
The gravitational potential for extended mass model can be evaluated as \cite{zakh07}:
\begin{equation}
\Phi_{ext}(r)=-G\int_r^\infty\frac{M(r^\prime)}{r^{\prime 2}}dr^\prime,
\label{equ05}
\end{equation}

We obtained the potential for extended mass distribution of matter in the form:
\begin{equation}
\begin{array} {lcl}
\Phi_{ext}(r) & = &-G\displaystyle\int\limits_r^{r_\infty}\dfrac{M_{ext}(r^\prime)}{r^{\prime 2}}dr^\prime = \\
& & \\
& = & \dfrac{{ - 4\pi {\rho _0}r_0^\alpha G}}{{\left( {3 - \alpha } \right) \left( {2 - \alpha } \right)}}\left( {{r_\infty}^{2 - \alpha } - {r^{2 - \alpha }}} \right),
\end{array}
\label{equ06}
\end{equation}

\noindent where $r_\infty$ is the outer radius for extended mass distribution of matter. The total gravitational potential can be obtained as a sum of Yukawa potential for central object with mass $M_{BH}$ (SMBH) and potential for extended matter with
mass $M_{ext}(r)$:
\begin{equation}
\Phi_{total}(r)=\Phi_Y(r)+\Phi_{ext}(r).
\label{equ07}
\end{equation}

Under assumption that the total potential $\Phi_{total}(r)$ does not differ significantly from Newtonian potential we obtain the
perturbed potential in the form:

\begin{equation}
V(r) = \Phi_{total} \left( r \right) - {\Phi_N}\left( r
\right); \quad {{\Phi_N}\left( r \right) =  - \dfrac{{GM}}{r}};
\label{equ08}
\end{equation}

\begin{equation}
V(r)=\Phi_Y(r)+\Phi_{ext}(r)-\Phi_N(r).
\label{equ09}
\end{equation}

Perturbing potential can be used for calculating the precession angle according to Eq. (30) in Ref. \cite{adki07}:

\begin{equation}
\Delta \theta = \dfrac{-2L}{GM e^2}\int\limits_{-1}^1 {\dfrac{z \cdot dz}{\sqrt{1 - z^2}}\dfrac{dV\left( z \right)}{dz}},
\label{equ10}
\end{equation}

\noindent where $r$ is related to the substitution variable $z$ via: $r = \dfrac{L}{1 + ez}$, and $L = a\left( {1 - {e^2}} \right)$ being the semilatus rectum of the orbit with semi-major axis $a$ and eccentricity $e$. By substituting the derivative of perturbing potential and $L$ in the above Eq. (\ref{equ10}), we obtain precession angle of S2 star orbit.

\section{Results and discussion}
\label{sec03}

We calculate orbital precession in Yukawa gravity potential and results are reported in Figs. \ref{fig01}--\ref{fig04} as a function of $\lambda$ and $\delta$. Taking the same values for orbital elements of S2 star in Yukawa gravity like in GR ($0^\circ.18$) we obtain numerically values for corresponding parameter $\lambda$.

Fig. \ref{fig01} shows the precession per orbital period for S2 star in ${\delta}-{\lambda}$ parameter space in the case of Yukawa modified gravity potential with the mass density distribution of extended matter $\rho_0$ = $2 \times 10^8 M_\odot \mathrm{pc^{-3}}$. Right panel represents enlarged part of the left panel. With a decreasing value of angle of precession, colors are darker. The obtained values of orbital precession angle in GR ($0^\circ.18$) is designated by dashed line. We can see
that parameter $\lambda$ that corresponds to orbital precession angle near $0^\circ.18$ is around 3100--4400 AU for small value of parameter $\delta$ (less than 10). For values of parameter $\delta$ between 10 and 40, dashed line tends to become vertical and depend very little on values of parameter $\delta$. The corresponding parameter $\lambda$ has values around 4400--4600 AU. For higher values of parameter $\delta$ (greater than 40), dashed line is vertical and does not depend on values of parameter $\delta$. The corresponding parameter $\lambda$ in this case has values around 4600 AU.

Fig. \ref{fig02} shows the same as in Fig. \ref{fig01}, but for the value of the mass density distribution of extended matter $\rho_0$ = $4 \times 10^8 M_\odot \mathrm{pc^{-3}}$. We can see the similar tendency regarding dependence of shape of dashed curve with respect to the parameter $\delta$, but now the parameter $\lambda$ has a smaller values with increasing of the mass density distribution of extended matter $\rho_0$. The parameter $\lambda$ that corresponds to orbital precession angle near $0^\circ.18$ is around 2000--3000 AU when $\delta$ is less than 10. For values of parameter ${\delta}$ between 10 and 40, the corresponding $\lambda$ has values around 3000--3100 AU. For values of parameter $\delta$ greater than 40, $\lambda$ has
values around 3175 AU.

Figs. \ref{fig03} and \ref{fig04} represent the same as Figs. \ref{fig01} and \ref{fig02}, but for the values of the mass density distribution of extended matter $\rho_0$ = $6 \times 10^8 M_\odot \mathrm{pc^{-3}}$, and $\rho_0$ = $8 \times 10^8 M_\odot \mathrm{pc^{-3}}$, respectively. We can see the similar tendency like in previous cases regarding dependence of shape of dashed curve with respect to the values of parameters $\delta$ and $\lambda$. We can conclude that changing of the mass density distribution of extended matter $\rho_0$ has strong influence on the precession per orbital period for S2 star.

Fig. \ref{fig05} shows comparison between the simulated orbits of S2-star in GR and in Yukawa gravity with the mass density distribution of  extended matter $\rho_0$ = $2 \times 10^8 M_\odot \mathrm{pc^{-3}}$ and Fig. \ref{fig06} shows the same like Fig. \ref{fig05}, but for the following values of parameters: $\lambda$ = 3000 AU and $\delta$ = 1. Simulated orbits in GR are obtained using parametrized post-Newtonian (PPN) equation of motion for two-body problem (for more details see \cite{zakh18a}).

The values of chosen parameters in Fig. \ref{fig05} are $\lambda$ = 3130 AU and $\delta$ = 1. For these values of parameters $\delta$ and $\lambda$ orbits obtained by GR and Yukawa model are very similar. If we change for instance (see Fig. \ref{fig06}) the parameter $\lambda$ = 3000 AU very little (less than 5 $\%$), difference between the orbits becomes very visible. That is why this method can help us to constrain modified gravity parameters.

\begin{table}[ht!]
\centering
\caption{The values of parameter $\lambda$ (in AU) for different combinations of 3 values of parameter $\delta$ the 5 values of the mass density distribution of extended matter $\rho_0$.}
\begin{tabular}{|l||c|c|c|c|c|}
\hline
& \multicolumn{5}{c|}{$\rho_0$ (in $10^8 M_\odot \mathrm{pc^{-3}}$)} \\
\cline{2-6}
& 0 & 2 & 4 & 6 & 8 \\
\hline
\hline
$\delta$=1  & 15125 & 3130 & 2080 & 1597 & 1302 \\
\hline
$\delta$=10 & 20395 & 4425 & 3015 & 2370 & 1978 \\
\hline
$\delta$=100 & 21285 & 4640 & 3175 & 2500 & 2090 \\
\hline
\end{tabular}
\label{tab01}
\end{table}

In Table \ref{tab01} we show the values of parameter $\lambda$, for different combinations of 3 values of parameter $\delta$ and 5 values of the mass density distribution of extended matter $\rho_0$. First column for $\rho_0$ represents the case when the mass density distribution of extended matter is not taken into account. These values are graphically represented in Figs. \ref{fig01}-\ref{fig04}. Clearly, from Table \ref{tab01} and Figs. \ref{fig01}-\ref{fig04} we can see tendency that if we fix precession angle to the value for GR ($0^\circ.18$), increasing of parameter $\delta$ strongly increase value of parameter $\lambda$ for $\delta$ between 0 and 10, but for higher values of $\delta$ increase of $\lambda$ is very small. On the other hand, increasing of the mass density distribution of extended matter $\rho_0$, decreases the value of parameter $\lambda$. Also, we can notice that if we do not take into account the mass density distribution of extended matter $\rho_0$ (first column), parameter $\lambda$ has much bigger values. It means that $\rho_0$ should be taken into account for calculation of the precession of S2 star orbit as well as other nearby stellar orbits which are close to SMBH.

\begin{table}[ht!]
\centering
\caption{The graviton mass ($m_g$) estimates corresponding to all mass density distributions presented in Table \ref{tab01}, in the case when Yukawa gravity parameter $\delta=1$.}
\begin{tabular}{|c||c|c|c|c|c|}
\hline
$\rho_0$ (in $10^8 M_\odot \mathrm{pc^{-3}}$)& 0 & 2 & 4 & 6 & 8 \\
\hline
\hline
$m_g$ (in $10^{-21}$ eV)  & 0.5 & 2.6 & 4.0 & 5.2 & 6.4 \\
\hline
\end{tabular}
\label{tab02}
\end{table}

Table \ref{tab02} contains the estimates for graviton mass $m_g$ in the case of all bulk distributions of matter from Table \ref{tab01} and for Yukawa gravity parameter $\delta=1$, which is usually considered in the massive gravity theory. These estimates are obtained according to: $m_g=h\,c/\lambda_g$, where $\lambda_g$ is the Compton wavelength of graviton which is assumed to be equal to the $\lambda$ parameter of Yukawa gravity \cite{zakh16,zakh18a}. By comparing these results for extended mass distribution with the estimates for upper bound on graviton mass from \cite{zakh16}, obtained without taking into account any bulk mass, one can see that our presented results still hold and are in expected range even for extended matter, but with lower mass densities $\rho_0 \lesssim 2 \times 10^8 M_\odot \mathrm{pc^{-3}}$.

\section{Conclusions}
\label{sec04}

We show that the mass density distribution of extended matter $\rho_0$ has significant influence on the value of precession angle per orbital period of S2 star. The parameter $\lambda$ has smaller values for larger mass density distributions of extended matter $\rho_0$. Therefore in these cases the corresponding estimates for graviton mass are slightly larger but stay in expected interval. A precession of orbit in Yukawa potential is in the same direction as in GR, but extended mass distribution produce a contribution to precession in opposite direction.

We can conclude that the mass density distribution of extended matter has significant influence on the values of precession angle and of modified gravity parameters. It means that mass density distribution of extended matter near the Center of SMBH should be taken into account when we want to calculate orbits of S stars.

\paragraph{Acknowledgments}
This work is supported by Ministry of Education, Science and Technological Development of the Republic of Serbia. P.J. wishes to acknowledge the support by this Ministry through the project contract No.  451-03-9/2021-14/200002.

\paragraph{Authors contributions}
All coauthors participated in calculation and discussion of obtained results. The authors contributed equally to this work.

\paragraph{Data Availability Statement}
This manuscript has no associated data or the data will not be deposited. [Authors' comment: All relevant data are in the paper.]

\end{document}